\documentclass[aps,prl,twocolumn,superscriptaddress,showpacs,draft]{revtex4}
\usepackage{graphicx}
\input{epsf}

\begin{document}

\title{Poincar\'e recurrences in Hamiltonian systems with 
a few degrees of freedom}

\author{D.L.Shepelyansky}
\affiliation{\mbox{Laboratoire de Physique Th\'eorique du CNRS (IRSAMC), 
Universit\'e de Toulouse, UPS, F-31062 Toulouse, France}}

\date{August 8, 2010; Revised: October 5, 2010}


\pacs{05.45.-a, 05.45.Ac, 05.45.Jn}
\begin{abstract}
Hundred twenty years after the fundamental work of Poincar\'e,
the statistics of Poincar\'e
recurrences in Hamiltonian systems with a few degrees of freedom
is studied by numerical simulations.
The obtained results show that in a regime, 
where the measure of stability islands
is significant, the decay of recurrences is characterized by a power
law at asymptotically large times. The exponent
of this decay is found to be $\beta \approx 1.3$. This value
is smaller compared to the average exponent $\beta \approx 1.5$
found previously for two-dimensional 
symplectic maps with divided phase space.
On the basis of previous and present results
a conjecture is put forward that, 
in a generic case with a finite measure of stability islands,
the Poncar\'e exponent 
has a universal average value $\beta \approx 1.3$
being independent of number of degrees of freedom
and chaos parameter. The detailed mechanisms of this slow 
algebraic decay
are still to be determined.
\end{abstract}

\maketitle

According to the Poincar\'e recurrence theorem proven in 1890
\cite{poincare1890} a dynamical trajectory with a fixed energy 
and bounded phase space will always return,
after a certain time, to a close vicinity of an initial state.
This famous result was obtained in relation to the studies of 
the three body gravitational problem
which fascinating history can be find in \cite{barrow}.
While recurrences will definitely take place a question about 
their properties, or what is a statistics of Poincar\'e recurrences,
still remains an unsolved problem. The two limiting cases of periodic or
fully chaotic motion are well understood: in the first case
recurrences are periodic while in the latter case
the probability of recurrences $P(t)$
with time being larger than $t$ drops exponentially
at $t \rightarrow \infty$ (see e.g. \cite{arnold,sinai}). 
The latter case is analogous to a coin flipping
where a probability to drop on one side after $t$
flips decays as $1/2^{t}$.

However, the statistics of  Poincar\'e recurrences
for generic two-dimensional (2D) symplectic maps
is much more rich. Such systems generally have a divided phase space
where islands of stable motion are surrounded by a chaotic component
\cite{chirikov1979,lichtenberg}. In such a case  trajectories
are sticking around stability islands and recurrences decay 
algebraically with time
\begin{equation}
\label{eq1}
P(t) \propto 1/t^{\beta} \; , \;\; \beta \approx 1.5 \; .
\end{equation}
The studies and discussions of this behavior can be find
in \cite{kiev,karney,chsh,ott,chirikov1999,ketzmerick,artuso}
and Refs. therein. According to the above studies the
Poincar\'e exponent $\beta$ has a universal average value
for 2D symplectic generic maps. 

While the statistics of Poincar\'e
recurrences in 2D maps has been studied in great detail 
\cite{kiev,karney,chsh,ott,chirikov1999,ketzmerick,artuso},
the original three body problem 
with a few degrees of freedom $N = 9$
addressed by Poincar\'e  \cite{poincare1890}
(effective number of degrees of freedom is $N_{eff}=6$
if to exclude the center of mass motion),
has not been studied yet in great detail.
The case of 4-dimensional and 6-dimensional symplectic maps
has been considered in \cite{bountis} and 
an algebraic decay of type (\ref{eq1})
has been found with $1.1 < \beta <1.5$
and  $1.7 < \beta < 2$ respectively.
A more detailed study, with up to $N=25$
degrees of freedom, has been performed in \cite{kantz}
with a variation of $\beta$ found to be in a range
$1.3 < \beta < 5.5$ depending on map parameters
and values of $N$.
In this work I study the statistics of Poincar\'e
recurrences in a model system for
$4 \leq N \leq 8$ going up to two orders of magnitude 
larger times comparing to \cite{bountis,kantz}.

\begin{figure}
\centerline{\epsfxsize=7.5cm\epsffile{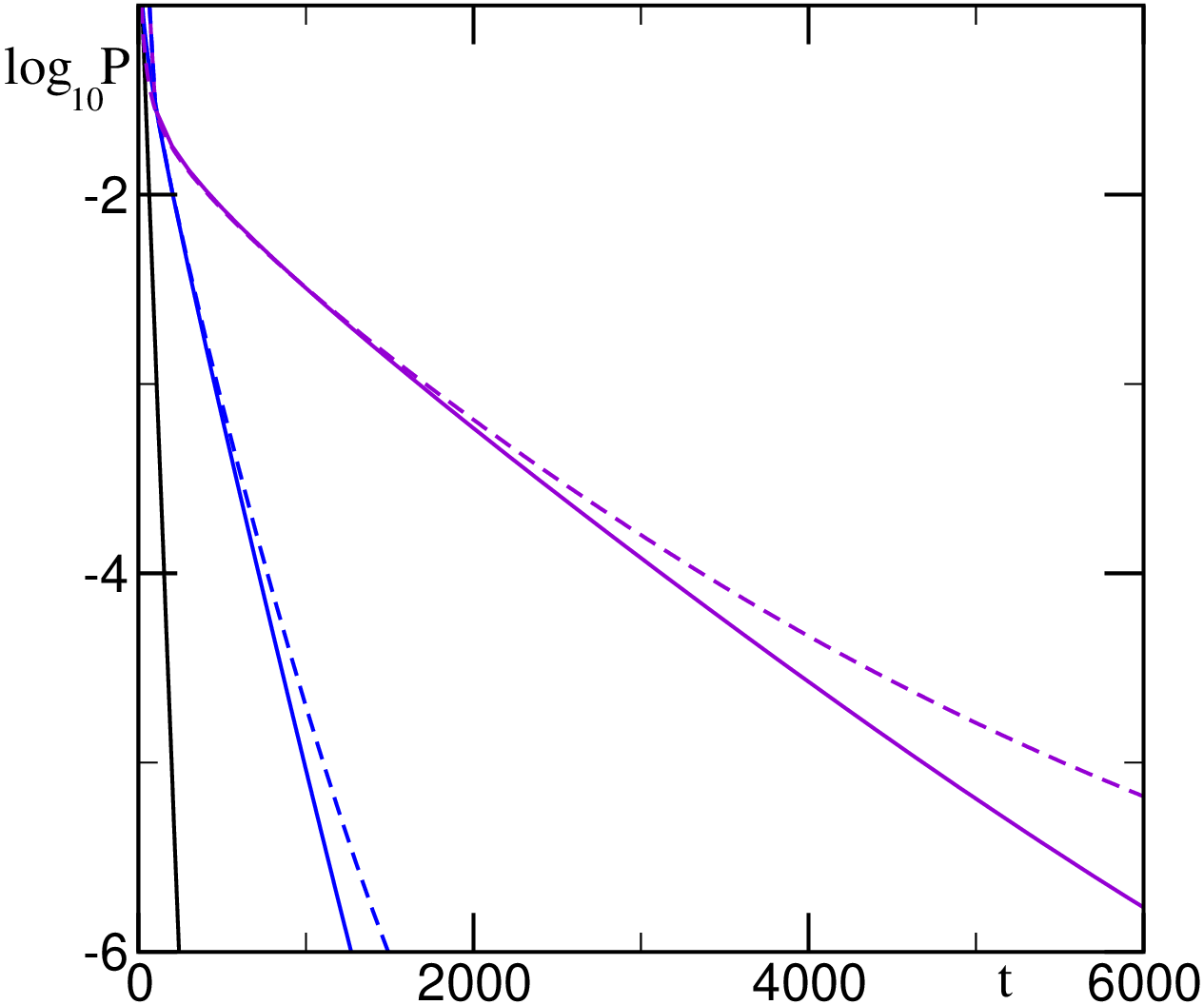}}
\vglue -0.2cm
\caption{(Color online) Dependence of statistics of Poincar\'e
recurrences $P(t)$ on time $t$ for 
$N=8$ and parameter $K=1, 0.6, 0.4 $ 
(full curves from left to right at $\log_{10} P =- 4$)
and for $N=6$ and $K=0.6, 0.4$ 
(dashed curves from left to right at $\log_{10} P =- 4$).
Here P(t) is an integrated probability of recurrences with time larger than $t$;
recurrences are considered on line $p_n=0$,
sum is taken over all $N$ degrees of freedom.
} 
\label{fig1}
\end{figure}

To reach a high efficiency of numerical simulations
I use  a dynamical map
\begin{eqnarray}
\label{eq2}
 {\bar p}_n  =  p_n + (K/2\pi) (\sin(2\pi(x_n - x_{n-1})) \nonumber\\
                             + \sin(2\pi(x_n-x_{n+1}))) \;, \nonumber\\
{\bar x}_n  =  x_n + {\bar p}_n \; ,
\end{eqnarray}
which was studied numerically in \cite{kaneko,chirikov1993,chirikov1997}.
Here bars mark new values of dynamical variables after one map iteration.
Periodic boundary conditions are used in $x_n ({\rm mod 1})$
and $p_n ({\rm mod 1})$ with $-0.5 \leq p_n \leq 0.5$.
The map is symplectic.
I use $N$ particles, $1 \leq n \leq N$, with a periodic boundary
conditions in $n ({\rm mod N})$. For $N=1$  the map (\ref{eq2})
is equivalent to the Chirikov standard map \cite{chirikov1979}
(assuming that all variables for $n>1$ are equal to zero).
The properties of $P(t)$ for this case can be find at
\cite{chsh,chirikov1999,ketzmerick,artuso} and Refs. therein.
For a number of particles $N>2$ the total momentum of 
the whole system is preserved so that
one can say that this situation corresponds effectively to
$N_{eff}=N-1/2$ degrees of freedom. In the following I consider
$4 \leq N \leq 8$.

The recurrences are considered on line $p_n=0$
for each particle, the integral probability
of recurrences, averaged over all particles, is defined as a total 
integral probability $P(t)$ of recurrences with
time larger than time $t$, which is measured in number of map iterations. 
In a more formal way, I count the number of map iterations $t_r$
between the consecutive crossing of line
$p_n=0$ for each particle, such an event is called a recurrence.
Then the relative number of recurrences with time $t_r$ larger than $t$
($t_r>t$) is taken to be equal to the recurrence probability $P(t)$
with averaging over all particles.

As in \cite{chsh,chirikov1999}, to compute $P(t)$ I usually used one trajectory iterated
up to time $t_{tot} \leq 10^{12}$. Special checks with other
trajectories or other $t_{tot}$ unsure that 
$P(t)$ remains unchanged in the limit of statistical
fluctuations which appear only when the number of recurrences
becomes of the order of a few events.
It should be noticed that the map (\ref{eq2})
is similar, in certain aspects, to the one studied in \cite{kantz}
(e.g. both are built on the basis of the Chirikov standard map),
but in the present case the couplings between particles are
local, while all particles are coupled in \cite{kantz}.

\begin{figure}
\centerline{\epsfxsize=7.5cm\epsffile{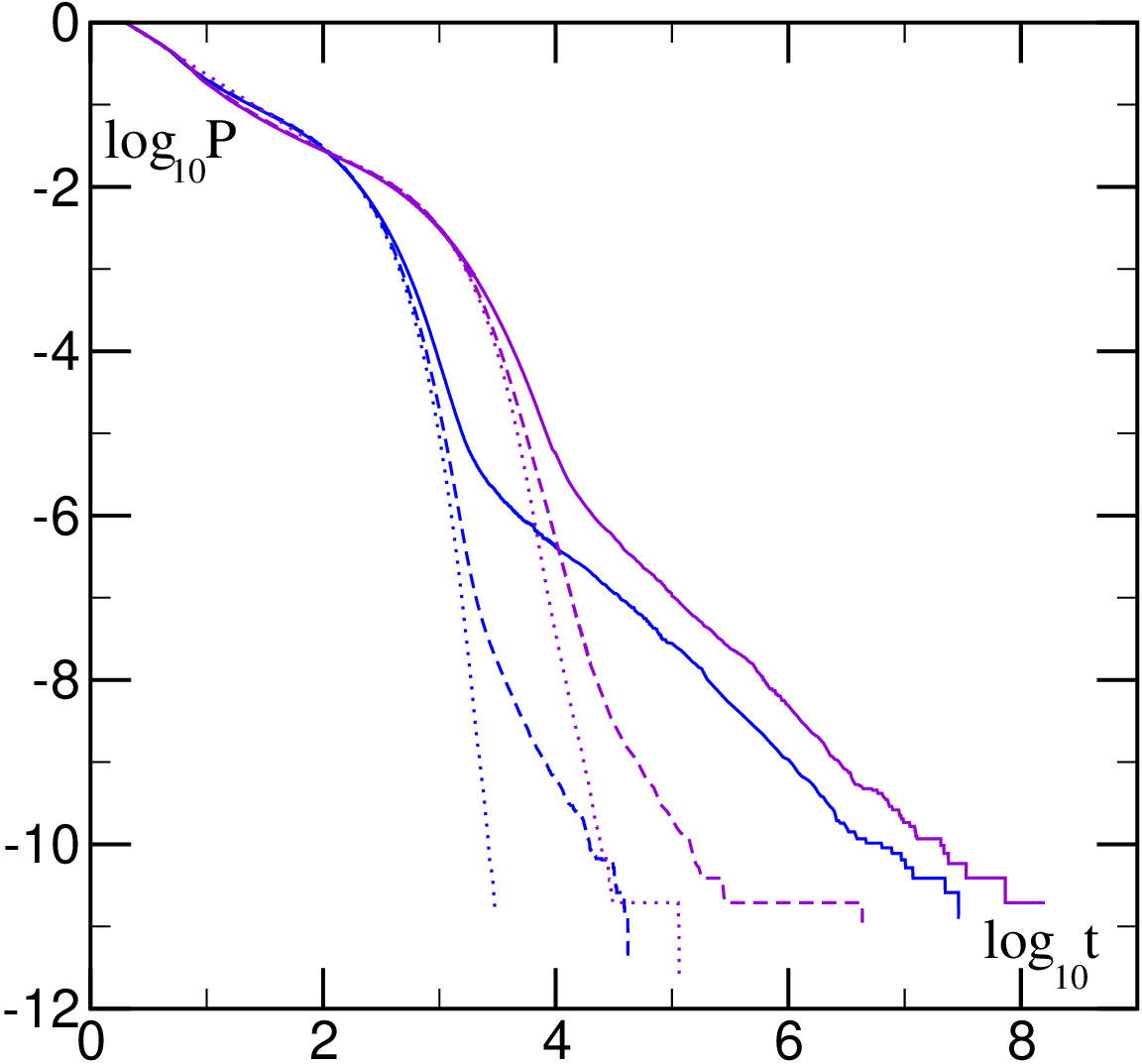}}
\vglue -0.2cm
\caption{(Color online) Same as in Fig.1  for  $K=0.6$ and
 $N=4, 6, 8$ (left group of blue/black full, dashed and dotted curves
from right to left at $\log_{10} P =- 8$ respectively)
and for $K=0.4$ and
 $N=4, 6, 8$ (right group of violet/gray full, dashed and dotted curves
from right to left at $\log_{10} P =- 8$ respectively).
The data are obtained from one trajectory with the
total number of iterations $t_{tot}=10^{12}$
(for $N=8$ I used $t_{tot}=10^{11}$).
} 
\label{fig2}
\end{figure}

An example of dependence of $P(t)$ on $t$ is shown in Fig.1
for relatively short times and large $N$ when the dynamics
is mainly fully chaotic. The initial decay drops exponentially
$P(t) \propto \exp(-t/t_D)$ with a certain time scale
$t_D$ which depends on $K$. The dependence of $t_D$ on
$N$ is relatively weak since up to a certain time
$P(t)$ curves are practically independent of $N$ 
(see Figs.~\ref{fig1},\ref{fig2}).
At large times the exponential decay is replaced by a power law
decay which is well visible for $N=4,6$ in Fig.~\ref{fig2}.
\begin{figure}
\centerline{\epsfxsize=7.5cm\epsffile{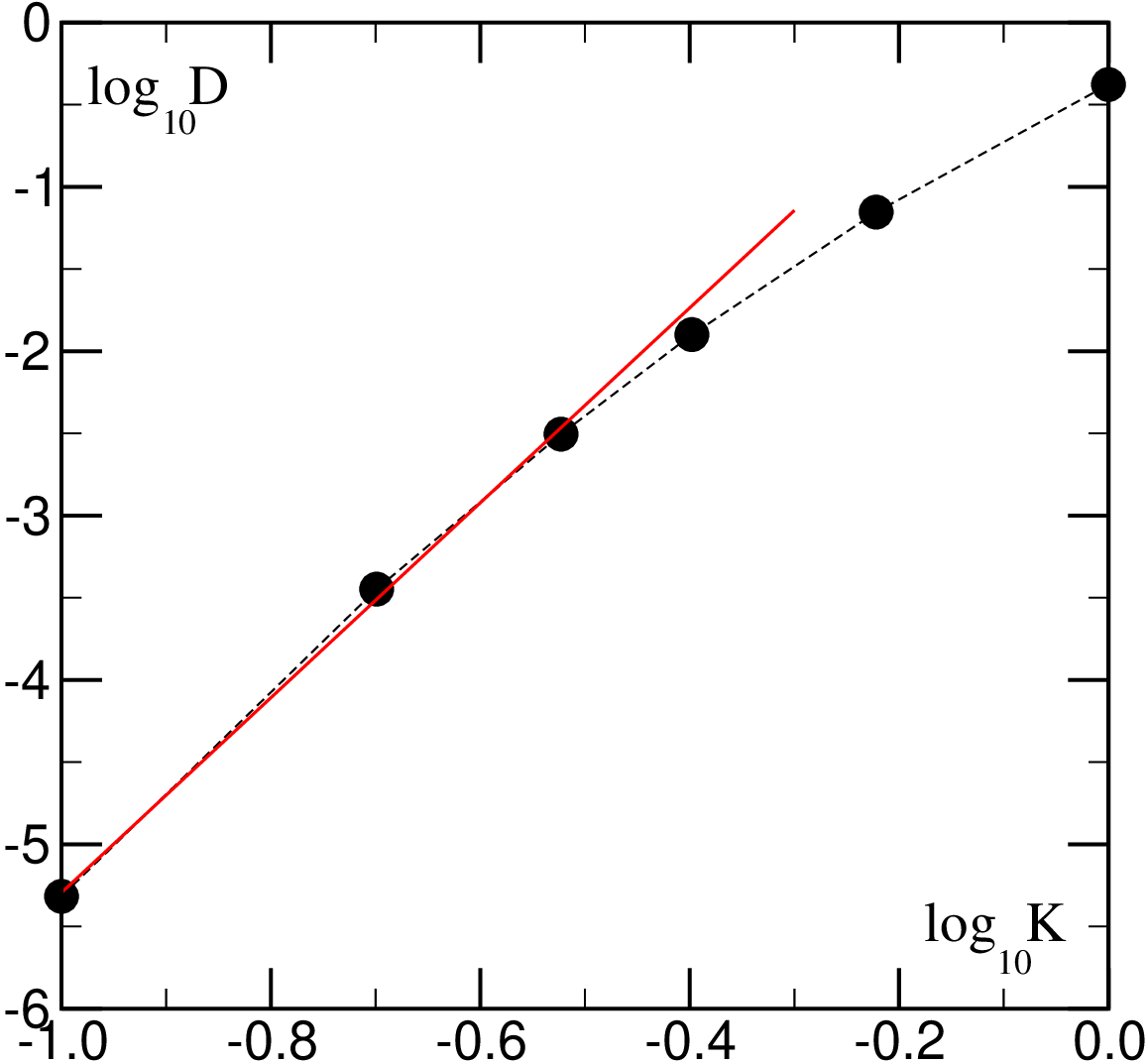}}
\vglue -0.2cm
\caption{(Color online) Dependence of the diffusion rate $D$
on chaos parameter $K$
(points). The dashed curve is drown to adapt an eye,
the full straight line shows the fit of last points
with $D=a K^b$ and $\log_{10} a =0.587$, $b=5.93 \pm 0.22$.
} 
\label{fig3}
\end{figure}

The time scale $t_D$ is related to a
diffusive spreading  in $p_n$ characterized by  a diffusion
rate $D/4\pi^2 =  <p_n^2>/t$. Indeed, such a relaxation diffusive
process on an interval $-0.5 \leq p_n \leq 0.5$ of size $L=1$
is described by the Fokker-Plank equation
\begin{equation}
\label{eq3}
\partial \rho /\partial t =D/(8\pi^2) \; \partial^2 \rho/\partial^2 p \;\; .
\end{equation}
This equation with zero boundary conditions $\rho(p=\pm 0.5)=0$
gives the exponential relaxation  of probability to stay inside 
the interval at
large times: $P(t) \sim \exp(-t/t_D)$ with 
$1/t_D = \pi^2 (D/4\pi^2)/(2L^2)=D/8$
(see e.g. Eq.(2.2.4) in \cite{redner}, in our case the interval size is $L=1$). 
Thus with this relation one can
extract from the initial
exponential drop of $P(t)$
the relaxation time $t_D$ and from it 
the diffusion rate $D$. In such a way 
I obtain the dependence of $D$ on $K$ and $N$.
As discussed above the dependence on $N$
is very weak and can be neglected.
On the contrary the dependence of $t_D$ and $D$ on $K$
is very strong as it is shown in Fig.~\ref{fig3}.

The dependence $D(K)$ has a few interesting features.
For $K=1$ I find $D \approx 1/2$ that corresponds to 
a random phase approximation valid in a regime of strong chaos.
With a decrease of $K$ the diffusion drops rapidly,
at small values of $K$ one has approximately algebraic decay
$D \propto K^b$ with the exponent $b=5.93 \pm 0.22$.
This value of the exponent is in a good agreement 
with the values obtained in \cite{kaneko,chirikov1997}
which are $b=6.6$ and $b=6.3$ respectively.
It should be stressed that the methods 
of computation of $D$ in \cite{kaneko,chirikov1997}
were rather different compared to those used here.

In fact an enormously powerful numerical method
has been used by Chirikov and Vecheslavov \cite{chirikov1997}
to compute an extremely small rate of the fast Arnold diffusion
(down to $D \sim 10^{-44}$ at $K \approx 8 \times 10^{-7}$
and $N=16$). This diffusion appears in very tiny chaotic 
layers around multi-dimensional resonances.
By its structure, the method used in \cite{chirikov1997}
determines the diffusion in a local domain of phase space while the method
used here gives the global diffusion. The agreement between two methods
shows that these two diffusion coefficient are
approximately the same.

\begin{figure}
\centerline{\epsfxsize=7.5cm\epsffile{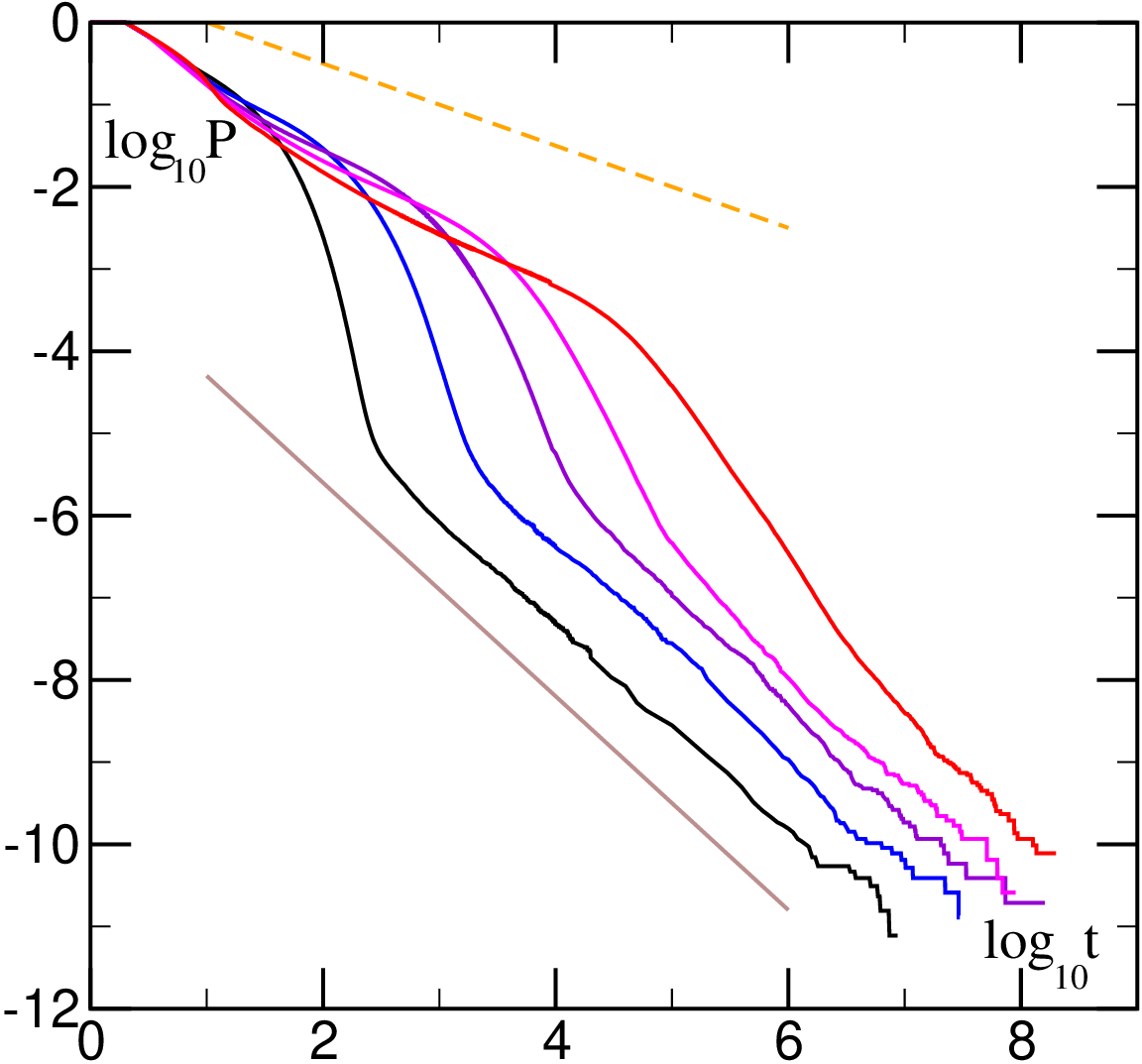}}
\vglue -0.2cm
\caption{(Color online) Statistics of Poincar\'e recurrences 
for the map (1) shown by curves for parameters
 $N=4$, $K=1, 0.6, 0.4, 0.3, 0.2$
(curves from left to right at $\log_{10} P =- 8$
respectively).
The exponents $\beta$ for the power law decay
$P(t) \propto 1/t^{\beta}$ are $1.243 \pm 0.001$, $1.292 \pm 0.002$,
$1.385 \pm 0.003$, $1.427 \pm 0.007$, $1.476 \pm 0.005$
respectively.
The full straight line shows the dependence
$P(t) \propto 1/t^{\beta}$ with 
$\beta=1.30 \pm 0.003$ corresponding to
the average of above 5 values of $\beta$.
The dashed straight line shows the diffusive decay
$P(t) \propto 1/\sqrt{t}$.
For each $K$ the data are obtained from one trajectory with the
total number of iterations $t_{tot}=10^{12}$.
} 
\label{fig4}
\end{figure}

In these studies I want to analyze how this chaotic web
influence the statistics of Poincar\'e recurrences.
Of course one is not able to go to so small values of $K$
but also in a certain sense one does not need this.
The algebraic decay of $P(t)$ appears due to sticking of trajectories around
stability islands so that one simply needs to have
a significant measure of stability islands.

The data of Fig.~\ref{fig2} show that for $N=8$ 
one has practically only an exponential decay of $P(t)$
indicating that the measure of stable component is 
of the order of $\mu_s \sim t P(t) < 10^{-8}$ for $K=0.6$
and $\mu_s < 10^{-5}$ for $K=0.4$
(I use the relation between $\mu$ and $P(t)$
discussed in \cite{chsh,chirikov1999}). For $N=6$ the algebraic decay
becomes to be visible at large $t$ showing that the measure of stability
islands starts to be reachable only for $t_{tot}=10^{12}$.

The power law decay of $P(t)$ is most visible for $N=4$ case
shown in Fig.~\ref{fig4}. Initially there is a slow decay
of $P(t)$ which is compatible with a diffusive spreading
on a semi-infinite line with $P(t) \propto 1/\sqrt{t}$
(see e.g. discussion at \cite{kiev}).
Since $t_D$ grows significantly with the decrease of $K$
the range of this diffusive decay of $P(t)$ increases
when $K \rightarrow 0$. However, already for
$K \leq 0.07$ the measure of chaotic component becomes
rather small and one needs to use special methods
described in \cite{chirikov1997} 
to be able to place initial conditions inside 
tiny chaotic layers. Due to these reasons I stop at 
values of $K \geq 0.1$. In any case 
for small $K$ the time $t_D$ becomes very large
and a lot of computational time becomes
lost for not very interesting diffusive decay.
 
 After the time scale $t_D$
a trajectory starts to feel a finite width of the chaotic layer 
with $-1/2 \leq p_n \leq 1/2$ and an algebraic decay due to sticking around
islands starts to be dominant. In this regime I find the exponent
$\beta =1.3$. The statistical error of this value is rather small
but certain oscillations in logarithmic scale of time 
are visible for $K=0.6, 0.4, 0.3$ so that the real uncertainty of
$\beta$ can be larger. 
At the same time  
the amplitude of these oscillations is significantly smaller
compared to the case of 2D symplectic maps discussed
in \cite{karney,chsh,chirikov1999,ketzmerick}.
The fit for $\beta$ is done  for times $t_{dr} < t < t_{tot}$ 
where $t_{dr}$ marks the end of the drop transition from
diffusive spreading to sticking in a  vicinity of islands.

The values of $\beta$, given in the caption of Fig.~\ref{fig4},
have a certain tendency to increase with a decrease of $K$.
However, this increase is rather small (about 19\%
while $K$ is changed by factor 5).
I attribute this to a decrease of fit interval 
at small values of $K$, where the diffusion time $t_D$
becomes larger and larger, that gives a reduction of 
the fit interval between $t_{dr}$ and $t_{tot}$.
It is clear that the fit interval 
$t_{dr} < t < t_{tot}$
for asymptotic algebraic decay should be sufficiently large
to determine $\beta$ reliably. This is clearly not so
for $N=6$ case shown  in Fig.~\ref{fig2}, 
where the transition from
exponential diffusive decay only starts to be replaced by 
an asymptotic algebraic decay. In my opinion a fit in such a small
interval would artificially increase the value of $\beta$,
since a sharp drop of $P(t)$ visible at $t<t_{dr} \sim t_D$
and being characteristic of diffusive exponential decay,
is not yet terminated completely. The data of Fig.~\ref{fig4}
clearly show that the scale $t_{dr}$ grows significantly
with a decrease of $K$ and $D$. 

This view, obtained on the basis of my results for 
rather long $t_{tot}$, leads me to another interpretation of previous
results \cite{bountis,kantz} which claimed the growth of $\beta$
with growth of $N$ and chaos parameter 
(see e.g. Fig.2 in \cite{kantz}).
Thus, on a first glance, in Fig.2(c) of \cite{kantz} $\beta$
increases from $\beta \approx 1.4$ to $2.8$ for $N=4$
when the chaos parameter $\xi$ is changed from $0.03$ to $0.1$.
This is in drastic contrast to the results presented here in Fig.~\ref{fig4}
clearly showing that $\beta \approx 1.3 \approx const$
when the chaos parameter is changed by a factor 5.
I think that such an increase of $\beta$ with $\xi$ in
\cite{kantz} should be attributed to shorter times 
considered there in comparison with the present studies.

In view of that I make a {\it conjecture} that 
in a generic case, when the islands of stability have nonzero measure,
{\it the asymptotic decay of Poincar\'e recurrences
has the form (\ref{eq1}) with a universal average
Poincar\'e exponent
$\beta \approx 1.3 - 1.4$ being independent of chaos parameter,
and number of degrees of freedom $N$ }(at least for 
moderate and large but finite
values of $N$).

The data of present studies confirm 
the approximate independence of $\beta$ of chaos parameter $K$
(see Fig.~\ref{fig4}). 
At the same time the data of Fig.2(c) of \cite{kantz}
at moderate values of chaos parameter $\xi=0.03$
clearly show that $\beta$ is approximately
$1.4 - 1.5$ for $2 \leq N \leq 10$. This confirms the above conjecture.
In my opinion, a further increase of $\beta$ for $11 \leq  N \leq 25$,
visible in Fig.2(c) of \cite{kantz} for $\xi=0.03$,
should be attributed to a significant reduction of the 
available fit interval $t_{dr} < t < t_{tot}$
which is clearly seen in Fig.2(a),(b)  of \cite{kantz}.
It is also clear that for the model of \cite{kantz}
the growth of $N$ gives an effective increase of the chaos parameter 
due to long range interactions present in the model. 
The data of \cite{bountis} for 4D map give approximately the same
universal value of $\beta$, while for 6D I expect that the time interval
was not so long to see the asymptotic behavior.

It is now well established that  generic 2D symplectic maps
have Poincar\'e recurrences with 
a universal average Poincar\'e exponent $\beta \approx 1.5$ 
\cite{kiev,karney,chsh,ott,ketzmerick}. 
This slow decay is linked to sticking in a vicinity of stability islands.
It is naturally to expect that for larger number of degrees of freedom
$N$ the structure of such sticking regions
is more complicated giving more possibilities for sticking
with slow Arnold diffusion processes.
Hence, intuitively it is natural to expect that 
for a few degrees of freedom the 
average value of $\beta$ will be smaller. The universal 
average value $\beta \approx 1.3 - 1.4$ found here and in \cite{kantz}
is in agreement with such expectations.

In conclusion, the studies of the statistics of Poincar\'e recurrences
in Hamiltonian systems with a few degrees of freedom show that
at large times it is characterized by a power law decay (\ref{eq1})
with the universal average exponent $\beta \approx 1.3$. 
This value is not so far
from the average exponent $\beta \approx 1.5$
found for the 2D symplectic maps. It is possible that
the physical mechanisms of this slow decay
have similar grounds related to sticking of trajectories
in a vicinity of small islands of stability for
enormously long times. 
Further extensive studies are required 
to understand in a deeper way  the detailed
mechanisms of this slow decay. Even more than hundred twenty years
after the work of Poincar\'e \cite{poincare1890}
this fundamental problem of dynamical chaos remains unsolved.

I thank A.S.~Pikovsky for stimulating discussions that
initiated this work.

\end{document}